\PassOptionsToPackage{table}{xcolor}
\documentclass[draft]{agujournal2019}
\usepackage{url} 
\usepackage[inline]{trackchanges} 
\usepackage{soul}
\usepackage{amsmath}
\usepackage{multirow}
\usepackage{tabularx}
\draftfalse

\journalname{JGR: Planets}

\begin{document}

\title{Electron-induced radiolysis of water ice and the buildup of oxygen}

\authors{Chantal Tinner\affil{1}, Andr\'e Galli\affil{1}, Fiona B\"ar\affil{1}, Antoine Pommerol\affil{1}, Martin Rubin\affil{1}, Audrey Vorburger\affil{1}, and Peter Wurz\affil{1}}

\affiliation{1}{Space Science and Planetology, Physics Institute, University of Bern, Bern, Switzerland}

\correspondingauthor{Andr\'e Galli}{andre.galli@unibe.ch}


\begin{keypoints}
\item Laboratory irradiation experiments are crucial to understand the surfaces of icy bodies in space
\item First mass spectrometer results from water ice regolith samples irradiated with electrons are shown
\item The dominant radiolysis products released from the ice samples are H$_2$ and O$_2$, with a fraction of the O$_2$ being retained in the ice 
\end{keypoints}

\begin{abstract}

Irradiation by energetic ions, electrons, and UV photons
induces sputtering and chemical processes (radiolysis) in the
surfaces of icy moons, comets, and icy grains. Laboratory experiments, both of ideal surfaces and of more complex and realistic analogue samples, are crucial to understand the interaction of surfaces of icy moons and comets with their space environment. This study shows the first results of mass spectrometry measurements from porous water ice regolith samples irradiated with electrons as a representative analogy to water-ice rich surfaces in the solar system. 

Previous studies have shown that most electron-induced H$_{2}$O
radiolysis products leave the ice as H$_{2}$ and O$_{2}$ and that O$_{2}$ can be trapped under certain conditions in the irradiated ice. Our new laboratory experiments confirm these findings. Moreover,
they quantify residence times and saturation levels of O$_{2}$ in originally pure water ice. H$_{2}$O may also be released from the water ice by irradiation, but the quantification of the released H$_{2}$O is more difficult and the total amount is sensitive to the electron flux and energy. 

\end{abstract}

\section*{Plain Language Summary}
The surface of any air-less body exposed to space is irradiated by charged and neutral particles. This irradiation can alter the surface and trigger chemical reactions (so-called radiolysis) and is an important factor in the context of icy satellites and their environment in the solar system. This paper presents one approach to better understand these irradiation processes by irradiating porous water ice samples with electrons at laboratory conditions representative of the icy moons of Jupiter. 

We find that most of the radiolysis products of water leave the ice as H$_{2}$ and O$_{2}$, while some of the produced O$_{2}$ remains trapped in the irradiated ice. This could help explain remote measurements of the surfaces of Europa and Ganymede. Water may also be released from the irradiated ice.

\section{Introduction} \label{sec:introduction}

Interest in plasma-induced radiolysis of water ice originates from observations in space. The irradiation of water ice surfaces by electrons, ions, and photons is a common process on many airless bodies in the solar system and beyond. Examples include the icy moons inside the magnetospheres of Jupiter and Saturn and small icy bodies and ice grains exposed to the solar wind. Understanding the irradiation processes of icy surfaces is also mandatory to correctly model the exospheres created by these processes \cite<see e.g.,>[]{Vorburger2022,Plainaki2018}.

The surface of Ganymede contains in the top layer 0.1--1\% by volume of O$_{2}$ relative to H$_{2}$O \cite{Calvin1996}. This estimate is based on the original detection of O$_{2}$ absorption lines at 577 and 627 nm by \citeA{Spencer1995}. This detection came as a surprise because pure solid O$_{2}$ would sublimate immediately at Ganymede surface temperatures \cite{Calvin1996}. It was suggested that the O$_2$ on Ganymede's trailing side is generated by magnetospheric bombardment \cite{Spencer1995,Calvin1996}. Absorption bands of solid O$_{2}$ were also observed on Europa's and Callisto's surfaces, albeit at lower abundance, implying an O$_{2}$/H$_{2}$O ratio of the order of 0.1\% \cite{Calvin1996}. Likewise, the presence of H$_{2}$O$_{2}$ on Europa's leading hemisphere was inferred by \citeA{Carlson1999} based on ultraviolet and infrared reflectance spectra. On the trailing hemisphere of Ganymede, the O$_{3}$ absorption band at 260 nm was measured with the Hubble Space Telescope \cite{Noll1996} and evidence for O$_3$ on Callisto was recently reported by \citeA{Ramachandran2024}, indicating that also O$_{3}$ exists in some irradiated icy surfaces of our solar system. The production of O$_{2}$ (and other oxidants) via radiolysis may also be relevant for the potential habitability of the subsurface oceans of Europa \cite{Johnson2003,Greenberg2010,Hand2020,Vance2023}. Efforts to reproduce observed abundances of ice radiolysis products such as O$_2$ in the surfaces and exospheres of icy moons are ongoing but are often hampered by an incomplete understanding of radiolysis processes in the regolith \cite<see, e.g.,>[] {Carberry2023}.

From previous laboratory experiments with electron-irradiated water ice (amorphous water ice films or granular, crystalline ice) at temperatures around $100\pm20$ K, it is known that the majority of radiolyzed water molecules leave the ice as H$_{2}$O $\rightarrow$ H$_{2} + \frac{1}{2}$ O$_{2}$ once irradiation has reached saturation levels \cite{Orlando2003,Petrik2006,Abdulgalil2017,Galli2018}. In colder water ice, electron irradiation also efficiently produces H$_{2}$O$_{2}$: 3\% of H$_2$O water molecules are radiolyzed into H$_{2}$O$_{2}$ at $T=12$ K according to \citeA{Zheng2006}, but a strong decrease in efficiency is observed for warmer ice temperatures \cite{Hand2011}. Regarding the relative abundances of released species, \citeA{Davis2021} measured that electrons ejected more H$_2$O than the radiolysis products H$_2$ and O$_2$ from water ice at temperatures below 80 K, whereas \citeA{Abdulgalil2017} and \citeA{Galli2018} measured considerably higher release rates of H$_2$ and O$_2$ than H$_2$O, albeit at ice temperatures above 90 K. \citeA{Petrik2005} observed that the released O$_2$ exceeded the D$_2$O for $T > 110$ K when they irradiated a D$_2$O ice film with electrons. One caveat to bear in mind when comparing different studies is that some authors use the term ``sputtering" to describe the sum of all electron-induced surface erosion processes, whereas electrons directly ejecting H$_2$O molecules from the ice surface is called ``electron-stimulated desorption" \cite{Petrik2005}.

Energetic ions impacting water ice can directly eject H$_2$O molecules in single collisions because of their larger mass compared to electrons, but ions can initiate radiolysis, too. Experiments show that ions predominantly sputter H$_2$O for water ice temperatures below 100 K, whereas the radiolysis product O$_2$ makes up a larger fraction of the total yield at increased ice temperatures \cite{Brown1984,Bahr2001,Baragiola2002,Teolis2009,Teolis2017}.
The laboratory experiment with argon irradiation of water ice by \citeA{Tribbett2021} shows that direct detection of O$_{2}$ is non-trivial but a crucial input to quantify the released O$_{2}$/H$_{2}$O ratios. The subsurface reservoir of O$_{2}$ is in any case crucial for Europa's exosphere since the O$_{2}$ column density inside Europa's irradiated surface layer is several orders of magnitude higher \cite{Hand2006,Li2022}) than the $4\times10^{14}$ cm$^{-2}$ in the atmosphere \cite{Johnson2003,Smyth2006,Roth2014,Li2020}.

Laboratory experiments are crucial to understand the interaction of the space environment with the surface of icy objects. Two complementary approaches exist: ideal-simplified and realistic-complex. The first approach is required to quantify individual release processes (such as sputtering or thermal desorption) and is usually performed on thin, compact ice films or even monolayers in ultra-high vacuum condition (e.g., \citeA{Baragiola2002,Petrik2005,Fama2008,Davis2021}). While such laboratory experiments are indispensable for a fundamental understanding of an individual process, an application gap between these results and surfaces of real celestial bodies in space may exist because of the co-occurrence of different processes on and in regolith ice. For instance, \citeA{Johnson2019} and \citeA{Carberry2023} argued for studying the fate of radiolysis products inside icy regolith to link the observed O$_2$ abundances and irradiation processes via exosphere models of Europa and Ganymede.
With the present laboratory experiments we therefore investigated how samples of water ice regolith react to irradiation for pressures and temperatures relevant for the icy moons of Jupiter. Porous ice consisting of ice grains in the size range of tens to hundreds micrometers and consisting of crystalline Ih (instead of amorphous) ice is representative of the surfaces of Ganymede and Europa \cite{Galli2016}.
Our study shows the first results of electron irradiation of these granular ice samples monitored with a newly developed mass spectrometer. We monitored the released species from electron-irradiated pure water ice at temperatures relevant for the Jovian icy moons, we watched out for effects of physical properties (grain size, crystallinity, and density) on the release of radiolysis products, we measured the release timescales of the major radiolysis products H$_{2}$ and O$_{2}$, and we compared these results from ice regolith samples with previous studies performed with ice films.
 
In this study, we will not attempt to derive absolute yields ({number of released molecules per impactor) from water ice upon electron irradiation. Different laboratory experiments and theoretical predictions agree that the yield ranges between a few times (0.1--1.0) molecules per electron for the most efficient electron energy of a few hundred eV for pure water ice \cite{Teolis2017,Galli2018,Davis2021}. For electron energies between 1--10 keV, the energy dependence of the yield appeared to differ between experiments with ice films \cite{Davis2021} and porous ice samples \cite{Galli2018}.

In Section~\ref{sec:methods}, we will briefly present our laboratory facilities, explain our measurement methods and ice sample production procedures and list the acquired data for this study. Section~\ref{sec:data_processing} will detail the data processing, followed by Section~\ref{sec:results} presenting the results. Section~\ref{sec:discussions} will discuss the implications of these results for icy moons and comets and Section~\ref{sec:conclusions} will conclude the paper. 

\section{Experimental Methods}\label{sec:methods}

We performed the irradiation experiments in the MEFISTO (MEsskammer für FlugzeitInStrumente und Time-Of-Flight) facility at the University of Bern described by \citeA{Galli2016}. The facility consists of a large vacuum chamber including a liquid-nitrogen-based cooling stage for the ice sample, mass spectrometers, and various options to irradiate samples with ions \cite<electron cyclotron resonance ion source,>[]{Marti2001}, electrons (electron gun, manufacturer: Kimball Physics), and UV light (broadband UV lamp and Ly-$\alpha$ available). For the current study, we focused on the electron irradiation of water ice samples at energies 0.5~keV to 5~keV. This energy range is representative for electrons in the Jovian bulk plasma or thermal plasma \cite{Liuzzo2020} that dominate the number flux of electrons onto Ganymede's poles \cite{Liuzzo2020} and for the low-energy end of the energetic electron distribution measured by the Galileo spacecraft \cite{Paranicas1999}.

In this study, we concentrated on the particles released from the irradiated samples and hence relied primarily on our custom-built time-of-flight mass spectrometer (TOF-MS) for monitoring the reactions introduced by irradiation. The TOF-MS follows the same design as the NIM mass spectrometer built for the Jupiter Icy Moons Explorer \cite{Foehn2021}. The TOF-MS accumulates mass spectra of all molecules from mass over charge 1--1000 inside the vacuum chamber at a time resolution of 0.1 s or multiples thereof. The achieved mass resolution varies depending on the mass peaks from 600 to 900 $m$/$\Delta m$, where $\Delta m$ is the full width at half amplitude of the peak. Where not otherwise specified, the masses will be rounded to the nearest integer mass over the charge $z = n\cdot e$, where $e$ is the elementary charge of an electron and $n$ is an integer number. Mass numbers 16~u/z and 18~u/z have average mass resolutions of:

\begin{equation}
  (m/\Delta m)_{16} = 680,  (m/\Delta m)_{18} = 750
\end{equation}

This mass resolution is sufficient to separate CH$_4$ from O, as depicted in Fig.~\ref{fig:mass res}, whereas other isobaric interferences cannot be resolved. A TOF mass spectrometer ionizes the species to accelerate them: Only ions can reach the detector. For simplicity, we will refer to the respective molecules in this work. The ions start an electron shower when hitting the microchannel plates, which is then converted to counts/s. When the yield is compared over various irradiations, the counts/s will be scaled to the irradiated area specific to the irradiation. We will indicate the measured intensities of released species in counts/s and scaled relative to the other species.
\begin{figure}[h]
\begin{center}
       \includegraphics[clip,width=0.65\textwidth]{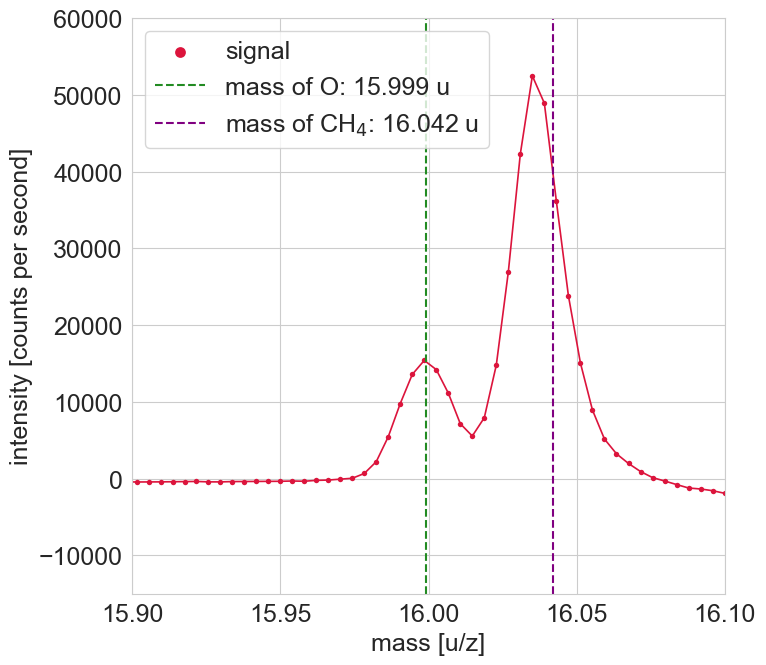}
        \caption{Extract from the data to examine the mass resolution of the time of flight mass spectrometer around 16~u/z. The red signal shows the measured data from which the averaged signal before the start of the irradiation was subtracted. The large peak likely includes a minor NH$_2$ contribution.}\label{fig:mass res}
\end{center}
\end{figure}

To create water ice samples, we relied on the Setup for Production of Icy Planetary Analogues (SPIPA) described by \citeA{Pommerol2019} to fabricate two types of porous ice samples, SPIPA-A and SPIPA-B. SPIPA-A ice consists of small ice grains at a low density (typical grain diameter 5 $\mu$m, bulk density $\approx$ 0.23 g cm$^{-3}$), whereas SPIPA-B ice consists of coarse grains at a higher bulk density (typical grain diameter 67 $\mu$m, bulk density $\approx$ 0.5 g cm$^{-3}$). Both ice sample types were produced from pure de-ionized water at ambient pressure (implying Ih crystal structure) and inserted in the sample holder on the pre-cooled cooling plate. For simplicity, we will call SPIPA-A samples ``fine-grained ice'' and SPIPA-B samples ``coarse-grained ice'' from here on. The aluminum sample holder has a cylindrical shaped cavity of 40 mm diameter and 9 mm depth. The thickness of the ice sample equals the sample holder's depth.
\begin{figure}[h]
    \centering
\includegraphics[trim=0cm 6cm 9cm 0cm,clip,width=0.99\textwidth]{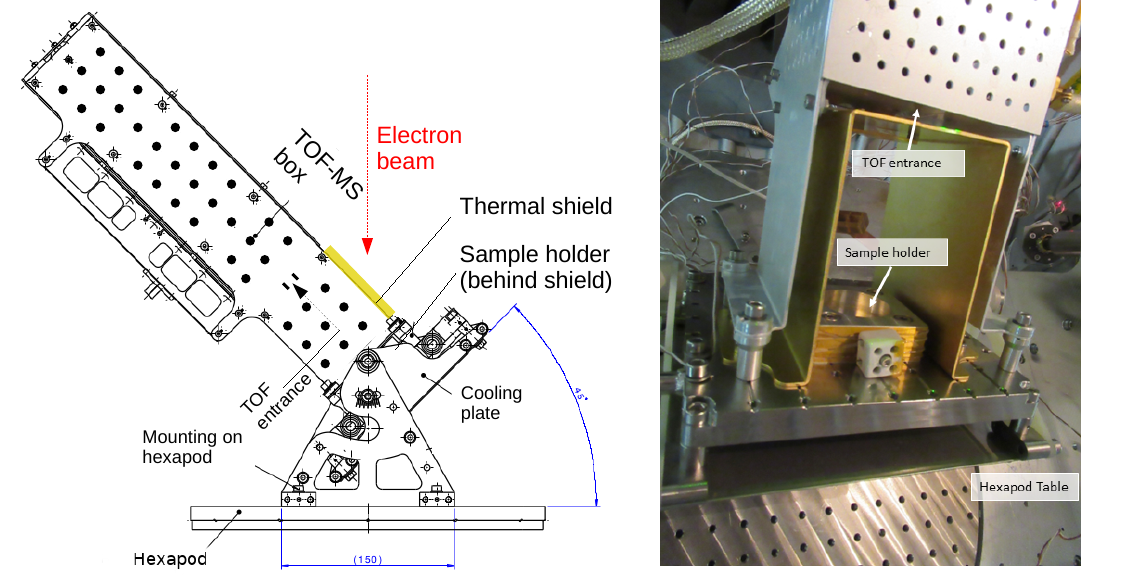}
     \caption{Experiment setup of the TOF-MS in the MEFISTO chamber. Left panel: Lateral view of the experiment setup with the TOF-MS box  mounted on the cooling plate. The electron beam from the top of the chamber impacts the ice in the sample holder at an incidence angle of $45^{\circ}$. Right panel: Photograph from the vantage point to the right of the sketch, showing the sample holder underneath the TOF-MS. The distance from the upper edge of the sample holder to the TOF-MS entrance is 100 mm.}\label{fig:sketch}
\end{figure}

The ice sample holder was mounted on a cooling plate centered below the TOF-MS, at a distance of 100 mm between the upper edge of the sample holder and the pin-hole (9 mm diameter) entrance to the TOF-MS. This is illustrated in Fig.~\ref{fig:sketch}. A gold-coated U-shaped thermal shield of dimensions 110 mm $\times$ 70 mm $\times$ 110 mm with a 10 mm hole at the TOF-MS entrance position was inserted between the ice sample and the MS (see Fig.~\ref{fig:sketch}) to minimize the heat transfer from the filament, the pulser, and the housing to the ice surface. The electrons impacted the ice at a $45^{\circ}$ incidence angle from the top of the chamber. The cooling plate and TOF-MS are placed on a movable hexapod table, which allows us to move the sample relative to the electron beam. A metal tape-covered copper ring directly above the sample holder rim (not shown in Fig.~\ref{fig:sketch}) acts as a Faraday cup measuring the electron beam intensity before and after irradiation. During all irradiation experiments, the ice sample holder was cooled with a constant flow of liquid nitrogen through the cooling plate. The temperature was monitored with several thermocouples on the sample holder and ranged between 91~K and 93~K. This is representative of the colder surface regions on the icy moons of Jupiter (minimum surface temperatures of roughly 80~K for Europa, Ganymede, and Callisto \cite{Orton1996,Moore2004,Ligier2019}). Crystalline ice produced from deionized water under equilibrium with normal laboratory atmosphere should contain less than 10 ppm of dissolved oxygen, which is orders of magnitude lower than the amount expected at the surface of the icy moons (see Section~\ref{sec:introduction}). These possible atmospheric contaminants should therefore not interfere with the amount of oxygen and other volatiles produced during irradiation and subsequent release.

After inserting the water ice sample into the sample holder, the chamber was sealed off and evacuated. Electron gun operations and TOF-MS measurements require ambient pressures below $10^{-6}$ mbar, which was usually reached after 12 hours of pumping. The typical ambient pressures during irradiation experiments were $(2\pm1)\times10^{-7}$~mbar. The residual gas consisted mostly of CO$_2$ and N$_2$ (see Section~\ref{sec:data_processing}) and did not notably affect the ice sample: no new or somehow different layer of ice or other contaminants of relevant thickness was formed on top of the ice sample even if 19 hours passed between two irradiations (see Section~\ref{sec:timescales}). We continued a series of experiments with a given ice sample for 2 or 3 days of measurements, irradiating different areas on the sample surface with an electron beam much narrower than the sample diameter of 40 mm. The electron beam was kept steady at one spot during irradiation, covering an oval or circular shaped area of 5--10 mm in diameter. These experiments were also designed to study if species release from pristine water ice differed in any way from the characteristics of previously irradiated water ice. An irradiation of an area that was exposed to the electron beam for the first time since the preparation of the sample was labeled ``pristine", in contrast to any experiment where a previously irradiated area (for a dedicated irradiation or any other test) of the ice sample was targeted again. The default time scale for an instantaneous release of a species upon irradiation is expected to be $\tau = V/S = 4.6$~s with the volume $V = 1.6$ m$^{3}$ and the pumping speed $S= 0.35$ m$^{3}$ s$^{-1}$ for MEFISTO \cite{Galli2016}.

We performed 24 irradiation experiments with electrons on pure water ice samples with irradiation times of 1 to 20 minutes and fluxes of $10^{13}$ to $10^{14}$ electrons cm$^{-2}$ s$^{-1}$. Among them, 15 irradiation experiments were found to release little or no observable water in the chamber and they coincided with cases of low electron beam currents or energies. These experiments are listed in Table~\ref{tab:Overview} plus two experiments where we irradiated the Faraday cup or the cooling plate above the sample holder instead of the water ice sample. The Faraday cup irradiation \#3 and cooling plate irradiation \#9 were included in the analysis as we suspected frost on these surfaces. 
The 9 other ice irradiation experiments were performed at a combination of higher electron fluxes and energies, which resulted in an unambiguous water release $\gg 10^5$ cnts~s$^{-1}$ (see bottom rows in Table~\ref{tab:Overview}). 

For the energies used in our experiment, the penetration depth $d$ [nm] of electrons in water ice \cite{Johnson1990,Hand2011} can be computed according to

\begin{equation}
    d =R_0 E^{\alpha},
    \label{eq:depth}
\end{equation}

and results in 13 -- 780~nm with $E$, the electron energy in units of keV, $\alpha =1.76$, and the depth $R_0= 46$~nm for samples with density $\rho$ = 1~g~cm$^{-3}$ at 1~keV. This is at least four orders of magnitude smaller than the thickness of our ice samples. This fact also holds true if we assume a deeper penetration depth because of the higher porosity of our ice samples relative to a compact ice film.
\\
\begin{table}[h]
\begin{tabular*} {\textwidth}{>{\footnotesize}l >{\footnotesize}r >{\footnotesize}r >{\footnotesize}r >{\footnotesize}r >{\footnotesize}r >{\footnotesize}r >{\footnotesize}r >{\footnotesize}r >{\footnotesize}r}

     & \textbf{Date} & \textbf{Pris-} & \textbf{Ice type} & \textbf{Flux} [10$^{13}$ & \textbf{Energy} & \textbf{Time } & \textbf{H$_{2}$ release} & \textbf{O$_{2}$ release} & \textbf{H$_{2}$O release} \\
    &  &  \textbf{tine?} &  & cm$^{-2}$s$^{-1}$] & {[}keV{]} & {[}min{]} &
    [$10^5$ cts s$^{-1}$] & [$10^5$ cts s$^{-1}$] & [$10^5$ cts s$^{-1}$] \\
    \hline
     \rowcolor{black!5}
     \#1 & 09.03., 15:07 & Yes & fine-gr. & 1.5 & 0.5 & 8 & $6.5\pm0.4$ & $2.7\pm1.3$ & $<1.0$ \\
   
    \#2 & 09.03., 15:26 & & fine-gr. &1.4 & 0.5 & 13 & $5.4\pm0.2$ & $0.5\pm0.3$ & $<0.1$  \\
    \rowcolor{black!5}
      \#3  & 09.03., 15:45 & & F-cup & 1.0 & 0.5 & 6 & $4.8\pm0.1$ & $0\pm0.1$ & $<0.05$  \\
   
    \#4  & 09.03., 17:40  & Yes & fine-gr. & 3.5 &0.5 & 14 & $18.3\pm0.5$ & $1.8\pm0.6$ & $<0.03$  \\
    \rowcolor{black!5}
      \#5  &10.03., 10:22& & fine-gr. & 0.8 & 1 & 9 & $3.6\pm0.4$ & $0\pm0.3$ & $<0.03$ \\
    
     \#6  & 10.03., 10:37 & & fine-gr. & 0.6  & 1 & 14 & $2.2\pm0.2$ & $0.3\pm0.1$ & $<0.2$  \\
     \rowcolor{black!5}
      \#7 & 10.03., 11:12& & fine-gr. & 2.2 &1 & 14 & $10.1\pm0.3$ & $2.4\pm0.2$ & $<0.03$ \\
   
    \#8 & 10.03., 11:37& & fine-gr. & 1.9 &1 & 3 & $8.7\pm0.2$ & $2.3\pm0.1$ & $<0.02$ \\
     \rowcolor{black!5}
      \#9 & 10.03., 11:48& & plate & 1.8& 1 & 1 & $9.8\pm0.4$ & $0.8\pm0.1$ & $<0.03$  \\
    
       \#10 & 10.03., 12:13 & Yes & fine-gr. & 7.3 & 1 & 14 & $33.1\pm2.1$ & $8.9\pm2.3$ & $0.27\pm0.13$ \\
     \rowcolor{black!5}
   \#11&  10.03., 13:11& & fine-gr. & 6.6 &1 & 21 & $25.3\pm3.6$ & $6.5\pm2.4$ & $<0.05$ \\
    
    \#12 & 10.03., 13:41&  & fine-gr. & 7.5 &1 & 3 & $25.9\pm1.1$ & $7.4\pm0.4$ & $<0.05$  \\
     \rowcolor{black!5}
    \#13 & 10.03., 14:18  & Yes & fine-gr. & 2.8 & 3 & 14 & $5.1\pm0.5$ & $0.7\pm0.4$ &$0.07\pm0.04$  \\
    
    \#14 & 16.03., 11:38 & Yes  & coarse-gr. & 0.6 &1 & 14 & $5.5\pm2.0$ & $2.6\pm0.9$ & $<0.1$ \\
     \rowcolor{black!5}
      \#15 & 16.03., 12:01 & & coarse-gr. & 0.6 &1 & 3 & $9.0\pm0.1$ & $2.5\pm0.1$ & $<0.1$  \\ 
    \#16 &  16.03., 12:13& & coarse-gr. &1.7 &1 & 3 & $22.4\pm0.3$ & $7.1\pm0.2$ & $0.33\pm0.27$  \\
    \rowcolor{black!5}
    \#17  &  16.03., 17:07 & Yes & coarse-gr. & 1.1 & 3 & 14 & $5.5\pm0.3$ & $0.5\pm0.6$ & $0.8\pm0.5$ \\ 
        \hline
     I & 10.03., 14:42& & fine-gr. & 14.5 & 3 & 9 & $70.7\pm4.3$ & $44.4\pm6.5$ & $273\pm131$\\
      \rowcolor{black!5}
      II & 11.03., 09:55  & Yes & fine-gr. & 1.1 & 5 & 14 & $6.0\pm1.0$ & $1.2\pm0.5$ & $5.7\pm3.5$ \\
    
      III &  11.03., 10:20 && fine-gr. & 1.5 & 5 & 14 & $8.2\pm1.5$ & $3.0\pm0.7$ & $8.0\pm5.4$\\
      \rowcolor{black!5}
      IV & 11.03., 10:46 & Yes &  fine-gr. &  1.7 & 5 & 19 & $7.9\pm1.4$ & $1.9\pm0.9$ & $3.9\pm2.1$ \\
     
     V & 11.03., 11:14& & fine-gr. & 2.5 & 5 & 3 & $15.7\pm1.4$ & $4.7\pm0.2$ & $18.8\pm3.3$  \\
      \rowcolor{black!5}
      VI & 16.03., 14:42& & coarse-gr. & 5.6 &1 & 3 & $66.6\pm2.8$ & $27.0\pm4.6$ & $14.4\pm7.5$\\
      
     VII & 16.03., 14:52& & coarse-gr. & 5.4 & 1 & 14 & $77.2\pm4.1$ & $31.3\pm3.7$ & $23.1\pm13.2$\\
      \rowcolor{black!5}
      VIII & 16.03., 17:30& & coarse-gr. & 2.4 &3 & 13 & $12.2\pm1.1$ & $4.3\pm1.4$ & $11.3\pm6.8$\\
    
      IX & 16.03., 17:54& & coarse-gr. & 10.7 &3 & 13 & $60.0\pm2.0$ & $29.9\pm7.6$ & $445\pm110$\\ 
\end{tabular*}

\caption{\label{tab:Overview} List of irradiation experiments, including information about the energies and flux of the electron beam, the type of ice sample (or the Faraday-cup or cooling plate in the case of test irradiations \#3 and \#9), and the integrated count rates (EICS and MCP gain corrections not included) of the main species once the release reached steady state. ``Pristine'' indicates that a hitherto un-irradiated area of the ice sample or a fresh ice sample was irradiated for the first time. Numbers I--IX list the experiments with a strong water release $>10^5$ cnts s$^{-1}$.}
\end{table}

\section{Data Processing}\label{sec:data_processing}

Before analyzing the gas species released from the irradiated ice sample, the stability of the signals before, during, and after the irradiation was assessed. To identify the net release of species during irradiation, the linear fits of the mass spectra accumulated before and after irradiation (the background levels) were subtracted from those during irradiation. This background signal should therefore be as stable as possible (no fluctuation due to pressure changes, movement of the hexapod table, or changes in the cooling system operation). Hydrocarbon species and their fragments were found to appear and disappear in the mass spectrum with the start and stop of the electron gun (see grey-red overview spectrum in Fig.~\ref{fig:subtracted_spectra}). This effect is most likely due to secondary electron desorption of contaminants off the chamber walls plus subsequent fragmentation to CH-fragments in the mass spectrometer. This hydrocarbon background also appears if the electron beam is directed at the (cold or warm) Faraday Cup or at another ice-free metal surface in the chamber. Here, we will only focus on the pure water ice radiolysis products from the ice sample.

The composition of the gas in the vacuum chamber was monitored before, during, and after each irradiation experiment at a default time resolution of 1 s. When the composition changed only slowly (during background measurements or irradiation when a new steady state was reached, integration times of 10 seconds were used.
An example of a TOF spectrum obtained during electron irradiation is shown by the grey line in Fig.~\ref{fig:subtracted_spectra}. The x-axis was converted from time-of-flight $t$ into mass per charge $m$ via the quadratic relationship $m = c_1(t-c_2)^2$ with two fit variables $c_1$ and $c_2$. This TOF spectrum is the sum of species released due to irradiation plus the background caused by residual gas. The latter is constant over the timescale of irradiations provided the cooling flow is constant. Thus, the background spectrum can be easily subtracted from the spectrum measured during irradiation, resulting in the net irradiation spectrum (shown in red in Fig.~\ref{fig:subtracted_spectra}). The species identified in these plots to react to ice sample irradiation were then further analyzed by means of time series of the signal strength as shown in Fig.~\ref{fig:timeseries_general}. These time series show the summed intensities in counts/s over the individual peaks of the masses of interest after subtracting the average spectrum baseline to the left and the right of the peak in the TOF spectrum.

The dominant background species in the chamber are CO$_2$ and N$_2$ (see Fig.~\ref{fig:timeseries_general}). They usually do not change when electron irradiation is started or stopped. These residual gas species thus would only pose a problem for analysis if the TOF-MS were at the limits of its dynamic range. The O$_2$ fragments from CO$_2$ in the mass spectrometer (pink line in Fig.~\ref{fig:timeseries_general}) overlap with radiolysed O$_2$, though, which is another reason why we use the changes relative to the background levels before and after irradiation to quantify radiolysis species.

\begin{figure}[htp]
  \includegraphics[width=0.95\textwidth]{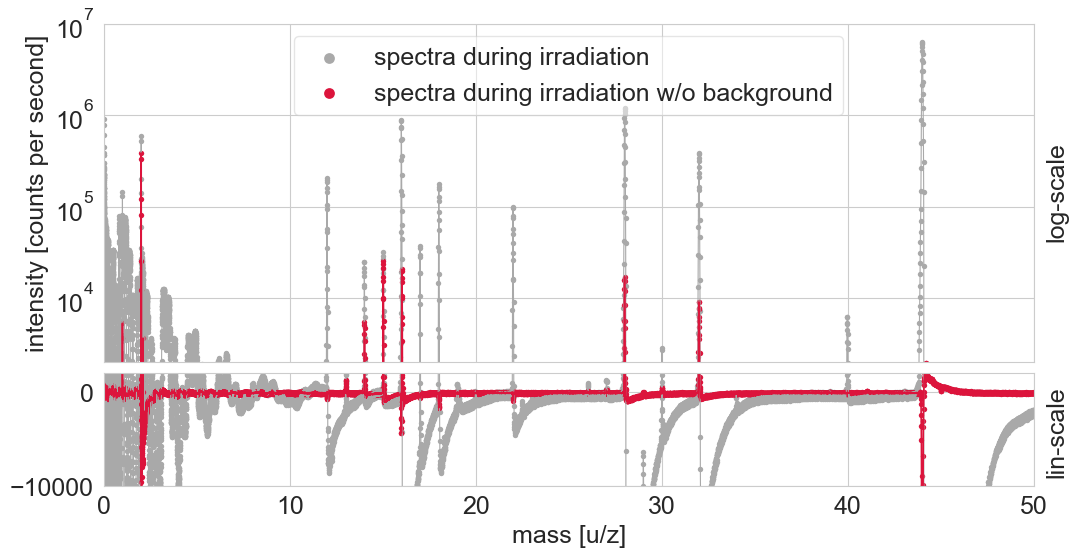}
\caption{Raw mass spectrum over irradiation time (grey symbols) and net mass spectrum (red symbols) after subtraction of the  background spectrum obtained before irradiation. The upper part of the plot is in logarithmic scale; the part underneath was kept linear to display the decreased sensitivity after strong peaks.}\label{fig:subtracted_spectra}
\end{figure} 

To derive the true relative abundance of a species from its measured signal strength in counts/s, the species-dependent sensitivity of the instrument must be considered. To this end, three different effects have to be taken into account. Microchannel plate (MCP) gain, species-dependent electron-impact ionization cross-section (EICS), and fragmentation of molecules. The MCP gain depends on molecular mass because lightweight species move at higher velocities for the same kinetic energy, causing a higher electron yield in the MCPs \cite{Meier1993}.
The detected signal has to be compensated for mass-dependent EICS, but here the larger molecules usually have a larger cross-section. Finally, ionization of molecules can lead to fragmentation. As a result, there always appears a signal at fragment species (e.g., HO) when the parent molecule (e.g., H$_2$O) is present. To reconstruct the initial parent-molecule abundance, one has to derive the intensity before fragmentation.

\citeA{Gasc2017} studied the fragmentation of various species with a Reflectron-type TOF-MS, similar to our set-up. The values he found for the fragmentation for water agree within 10 \% with the ones from \citeA{NIST_spectra} whose data are based on measurements from a quadrupole mass spectrometer. Therefore we used the fragmentation probabilities from \citeA{NIST_spectra} listed in Tab.~\ref{tab:FRAG_and_ISO}.

\begin{table}[!ht]
\begin{tabular}{lllllll}
\textbf{Species  }   & \textbf{Mechanism}   & \textbf{Product  }   & \textbf{Abundance}      & \textbf{Relative to}                                               \\
\hline
\multirow{4}{*}{H$_2$O}  & \multirow{2}{*}{Fragments to}    & O                      & 0.0090 $^{\text{{[}a{]}}}$             & H$_2$O              & \multicolumn{2}{l}{}            \\
                      &                                   & OH                     & 0.2121 $^{\text{{[}a{]}}}$             & H$_2$O              & \multicolumn{2}{l}{}                               \\
           
                      & \multirow{2}{*}{Isobars} & $^{17}$OH & 0.00038   $^{\text{{[}d{]}}}$            & OH               & \multicolumn{2}{l}{}                                                                                \\
                      &                                   & $^{18}$O  & 0.0020 $^{\text{{[}d{]}}}$              & O                & \multicolumn{2}{l}{}                                 \\
                       \hline
                      
\multirow{2}{*}{H$_2$}   & Fragments to                      & H                      & 0.0210 $^{\text{{[}a{]}}}$             & H$_2$               & \multicolumn{2}{l}{}             \\

                      & Isobars                  & D                      & 0.000156 $^{\text{{[}c{]}}}$                 & H                & \multicolumn{2}{l}{}                                 \\
\hline
O$_2$                    & Fragments to                      & O                      & 0.2180 $^{\text{{[}a{]}}}$             & O$_2$               & \multicolumn{2}{l}{}             \\ \hline 
\multirow{4}{*}{H$_3$O} 
                      & \multirow{4}{*}{Isobars} & HDO                    & 0.0003 $^{\text{{[}c{]}}}$             & H$_2$O              & \multicolumn{2}{l}{}                                 \\
                      &                                   & H$^{18}$O                   & 0.0020 $^{\text{{[}d{]}}}$             & OH               & \multicolumn{2}{l}{}                                 \\
                      &                                   & H$_2$~$^{17}$O                  & 0.00038 $^{\text{{[}d{]}}}$            & H$_2$O              & \multicolumn{2}{l}{}                                 \\
                      &                                   & D$^{17}$O                   & $<$ 0.0001 $^{\text{{[}d{]}}}$  & OH               & \multicolumn{2}{l}{}                                 \\
                      \hline
\multirow{2}{*}{OH}   & Receives fragments                & H$_2$O                    & $0.245\pm0.006$ $^{\text{{[}g{]}}}$             & H$_2$O               \\
                      & Isobars                  & $^{17}$O                    & 0.00038 $^{\text{{[}d{]}}}$            & O                & \multicolumn{2}{l}{}                                 \\
                      \hline
\multirow{3}{*}{H$_2$O$_2$} & \multirow{3}{*}{Isobars} & $^{16}$O$^{18}$O                 & 0.00401 $^{\text{{[}d{]}}}$            & O$_2$               & \multicolumn{2}{l}{}               \\
                      &                                   &  $^{17}$O$^{17}$O                 &$<$ 0.0001 $^{\text{{[}d{]}}}$  &  O$_2$                & \multicolumn{2}{l}{}                                 \\
                      &                                   & DO$_2$                    & 0.000156 $^{\text{{[}b{]}}}$ $^{\text{{[}c{]}}}$  &   HO$_2$               & \multicolumn{2}{l}{}                                 \\
\multirow{3}{*}{} & \multirow{3}{*}{Fragments to} & HO$_{2}$                 & 0.087 $^{\text{{[}e{]}}}$            & H$_2$O$_2$               & \multicolumn{2}{l}{}              \\
                      &                                   &  OH                 & 0.198 $^{\text{{[}e{]}}}$  &  H$_2$O$_2$                & \multicolumn{2}{l}{}                                 \\
                      &                                   & O                    & 0.02 $^{\text{{[}e{]}}}$  &   H$_2$O$_2$              & \multicolumn{2}{l}{}                                \\
                      \hline
\multirow{2}{*}{O}    & Receives fragments                & O$_2$                     & 0.2180 $^{\text{{[}a{]}}}$             & O$_2$               & \multicolumn{2}{l}{}           \\
                      & Isobars                  & CH$_4$                    & -                         &                  & \multicolumn{2}{l}{}                                   \\
                      \hline
H                     & Receives fragments                & H$_2$                     & 0.0210 $^{\text{{[}a{]}}}$             & H$_2$               & \multicolumn{2}{l}{}          \\    \hline
CO$_2$    & Fragments to                & O                     & 0.0961 $^{\text{{[}a{]}}}$             & CO$_2$         \\ \hline
O$_3$ & \multirow{2}{*}{Fragments to} & O$_2$& 5.0 $^{\text{{[}f{]}}}$  & O$_3$ & \multicolumn{2}{l}{} \\
& & O &  0.465 $^{\text{{[}f{]}}}$ & O$_3$ & \multicolumn{2}{l}{}                                                                                                        \\
\end{tabular}
\caption{\label{tab:FRAG_and_ISO} Table of the used fragmentation and isotopes ratios. The fragmentation patterns [a] are from \citeA{NIST_spectra} and the isotope data are from \citeA{NIST_isotopes_masses} [b], from \citeA{hydrogen} [c], and from \citeA{oxygen} [d], the hydrogen peroxide fragmentation data were taken from \citeA{Lindeman1959} [e], the ozone fragmentation data were taken from \citeA{Herron_1956} [f], and the OH/H$_2$O ratio was measured with our setup [g]. For the abundant species H$_2$O and OH, deuterium-bearing isobars were neglected as they would change the inferred relative abundances by only $\sim$ $10^{-4}.$ While the `Fragments to' is useful mostly to reconstruct the total peak height before fragmentation, the `Receives fragments' is essential as major species (only mentioned here: H$_2$, O$_2$, and H$_2$O) fragment to less abundant ones. The fragments are relative to 100 \% of the fragmenting species.}
\end{table}

According to \citeA{Gasc2015}, the electron ionization cross-section can be used as a first-order spectral deconvolution.
The model used by \citeA{Kim2005} to calculate the electron impact cross-sections is the binary encounter Bethe model for electron-impact ionization. According to the Bethe model, the ionization cross-section depends on various parameters, but none differ between isotopes. From an experimental standpoint, \citeA{Itikawa2005} performed experiments to compare the ionization cross-sections of various molecules containing hydrogen using deuterium molecules. From this, we conclude that there is a consensus that the effective cross-sections for isotopes are the same. 

The microchannel plate gains were investigated by \citeA{Meier1993}. They depend on the voltages applied to the microchannel plates. In our mass spectrometer, an experiment with test gas resulted in a gain for H$_2$ and He of 1.8 relative to N$_2$ (assuming the EICS as seen in Tab.~\ref{tab:EICS and MCP}). On the other hand, krypton had a similar gain as N$_2$. We, therefore, scaled only the lighter species (H, H$_2$ and He) with 1.8. The table with the values for EICS and MCP is shown in Tab.~\ref{tab:EICS and MCP}. Unless mentioned otherwise, unscaled values are used throughout the paper. Where the count rates are compensated for MCP gain, EICS, and fragmentation, they are scaled relative to N$_2$.

\begin{table}[h]
    \centering
\begin{tabular}{lll}
\textbf{Species} & \textbf{EICS {[}$\text{\AA}^2${]}} & \textbf{MCP gain} \\
\hline
H & 0.601     & 1.8                    \\
H$_2$ & 1.021  & 1.8                   \\
O & 1.363 & 1 \\
H$_2$O & 2.275 & 1                     \\
O$_2$ & 2.441 & 1                   \\
N$_2$ & 2.508   & 1       \\
O$_3$ & 3.520 & 1
\end{tabular}
    \caption{\label{tab:EICS and MCP} Electron impact cross-sections for ionization at 70~eV and microchannel plate gains for selected species. Data were taken from NIST \cite{Ionization}. The MCP gains are normed relative to N$_2$.}
\end{table}
\section{Results}\label{sec:results}

\subsection{Detected Species}\label{sec:relevant_species}

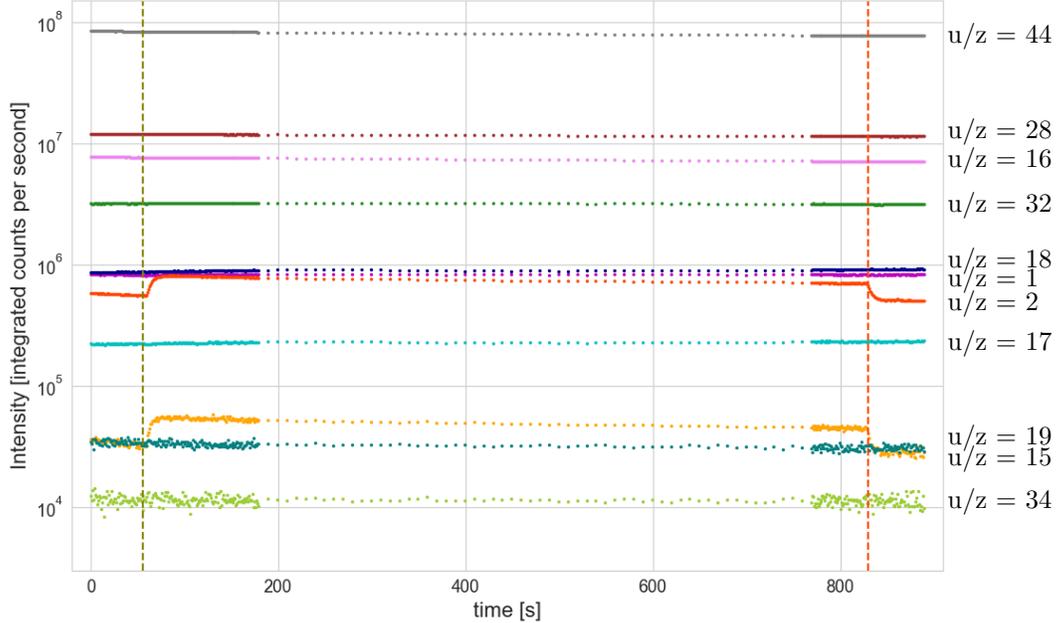
\begin{figure}[h]
  \includegraphics[trim=0cm 0cm 6.2cm 0cm,clip,width=0.9\textwidth]{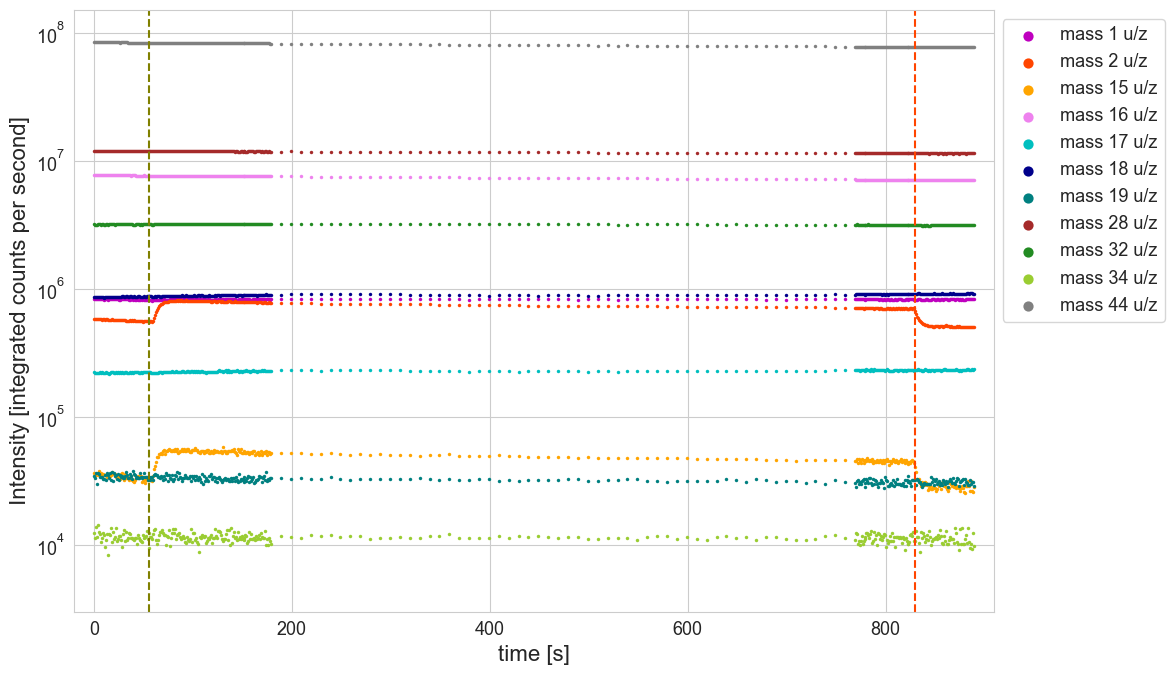}\put(0,222){u/z = 44}\put(0,186){u/z = 28}\put(0,175){u/z = 16}\put(0,158){u/z = 32}\put(0,137){u/z = 18}\put(0,130){u/z = 1}\put(0,121){u/z = 2}\put(0,106){u/z = 17}\put(0,70){u/z = 19}\put(0,62){u/z = 15}\put(0,46){u/z = 34} 
\caption{Time series of relevant species for irradiation experiment \#7. The intensities are the integrated counts per second over a variable mass width depending on the width of the peaks. The MCP and EICS gains were not applied here. The start and end of the irradiation are marked with a green and red dashed line, respectively. The intermediate data points between 180 s and 780 s were sampled at a lower time resolution of 10~s instead of 1~s. For a zoom-in on the H$_2$ (red) and O$_2$ (dark green) time series, see Fig.~\ref{fig:fitted_data}.}\label{fig:timeseries_general}
\end{figure}

Among the expected water ice radiolysis and sputtering products, H$_2$, O$_2$, and (depending on electron flux and energy) H$_2$O were observed to rise and fall with electron irradiation of the ice. Other possible radiolysis species such as O$_3$, H$_2$O$_2$, and OH were not identified unambiguously so far, taking into account isotope abundances of major species such as $^{16}$O$^{18}$O at 34~u/z.

Figure~\ref{fig:subtracted_spectra} shows a typical mass spectrum before (grey circles) and after background subtraction (red circles). The background-subtracted spectra were used to identify which species were released by electron irradiation. These species can be divided into different categories: First, there are species related to the radiolysis or sputtering of water molecules: 2~u/z (H$_2$), 32~u/z (O$_2$) with the expected fragmentation patterns at 1~u/z (H) and 16~u/z (O), respectively. Other water-related species could appear, e.g., at 17~u/z (HO), 18~u/z (H$_2$O), and 19~u/z (H$_3$O). However, these are not evident for the specific case in Fig.~\ref{fig:subtracted_spectra}.
We also notice a strong, constant signal at 44~u/z, CO$_2$ and double ionized CO$_2$ at 22~u/z, and some rising species not related to water ice: hydrocarbon chains at masses 12, 13, 14, 15, and 28~u/z. The signal around 16~u/z, can be resolved in a double-peak attributable to atomic oxygen at 15.99 and methane at 16.03, whereby the methane contributes more strongly to the rise seen as a result of the electron gun switch-on. These hydrocarbon species at 12, 13, 14, 15, 16, and 28 are not produced in ice radiolysis but rather the fragments of larger chains of hydrocarbons desorbed from chamber surfaces (see Section~\ref{sec:data_processing}). Figure~\ref{fig:timeseries_general} shows a typical time series (for irradiation \#7) of the identified species released to the chamber. The dashed lines indicate the start and the end of the electron irradiation.

\subsection{H$_2$O Release: Thermal versus Cold Trap Effects} \label{sec:h2o}

Most irradiation experiments showed little or no H$_2$O release if the product of electron flux and energy was low, whereas a strong water signal (often dominating over H$_2$ and O$_2$) was observed once flux times energy reached $10^{14}$~keV~cm$^{-2}$~s$^{-1}$ (for fine-grained ice) or $5 \times 10^{13}$ keV~m$^{-2}$~s$^{-1}$ (for coarse-grained ice). This is illustrated in Fig.~\ref{fig:H2O_prod}. Moreover, the observed water release for irradiations above these thresholds rose in a significantly slower and more irregular manner compared to the simple exponential increase of H$_2$ and O$_2$ reaching saturation levels after a few seconds (in the case of pre-irradiated ice samples).

\begin{figure}[h]
  \includegraphics[width=0.95\textwidth]{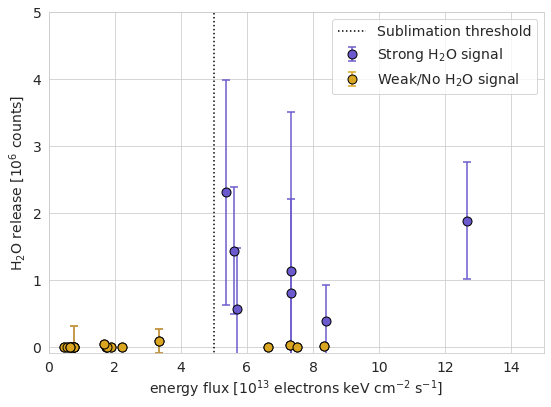}
\caption{Water release (steady state) as a function of electron energy flux for irradiation experiments with little or no discernible water release (orange symbols, experiments \#1--\#17 without \#3 and \#9 in Table~\ref{tab:Overview}) and for experiments with strong water signals (blue symbols, experiments I--IX in Table~\ref{tab:Overview}, the two data points with water release $>10^7$ cnts s$^{-1}$ were omitted). Above a threshold of $5\times10^{13}$ keV cm$^{-2}$ s$^{-1}$, intense H$_2$O signals can be released from the ice sample.}\label{fig:H2O_prod}
\end{figure}

The rapid increase of H$_2$O once the electron energy flux exceeds a limit (see Fig.~\ref{fig:H2O_prod}) can be understood as a thermal effect: The energy deposited by the electron gun warms up the top surface layers of the porous regolith, because of the very low thermal conductivity. This heating up is an experimental challenge, which does not come into play on the surfaces of icy moons: in the laboratory, one has to use electron fluxes that are many orders of magnitude higher, typically $10^{13}$ electrons cm$^{-2}$ s$^{-1}$ for the experiments described here or by \citeA{Davis2021}, than in space to reach saturation levels and/or measurable signals within minutes or hours. However, the surface of the ice sample may warm up to a temperature notably higher than the controlled sample holder as a consequence (see Fig.~\ref{fig:H2O_prod}).

While this effect tends to overestimate the H$_2$O release from electron irradiation from icy surfaces in outer space, the cold trapping of water molecules on cold surfaces inside the chamber (see right panel in Fig.~\ref{fig:sketch}) may lead to an underestimation:
Due to its high sticking efficiency \cite{Gibson2011}, water released from the ice may be under-represented relative to more volatile species such as H$_2$ and O$_2$ \cite{Bar-Nun1985,Cassidy2005,He2016,Davis2021}. In our setup this underestimation could reach up to a factor of 100, judging from the observation that the O$_2$/H$_2$O ratio for experiments with high electron fluxes (where the ice sample starts sublimating as a whole) is on the order of 0.1--1 whereas the unbiased O$_2$/H$_2$O bulk abundance should be  0.001--0.01 if our samples are a realistic analogue to granular icy surfaces on Europa and Ganymede \cite{Spencer2002,Hand2006}.

Because of the thermal and cold trapping effects, we do not attempt to derive a H$_2$O/O$_2$ release ratio in this analysis. One possible strategy for future experiments to observe an unbiased ratio would be to cool down all walls surrounding the sample to a temperature where also most other volatiles (in particular H$_2$ and O$_2$) would freeze out. However, this would require a major design change of the experiment setup, including a He-cryofinger. The potential under-representation of H$_2$O will therefore first be assessed by varying the observation geometry, i.e., the distance and viewing angle between the irradiated ice sample and the TOF-MS entrance. The sublimation and cold trapping effects do not affect the main radiolysis products H$_2$ and O$_2$.

\subsection{H$_2$ and O$_2$ Release}\label{sec:timescales}

H$_2$ and O$_2$ are the released species identified in most electron irradiation experiments performed here.
The ratio of H$_2$ to O$_2$ is a practical way of testing the experimental setup (particularly regarding the presence of background species not related to ice radiolysis), as its theoretical value is well known. When irradiating water ice, one expects a steady state ratio $r$ of 2:1 of released H$_2$ to O$_2$ if the water molecules fragment and there is no detection bias from the experimental conditions.

To interpret the measured ratio one has to take into account the MCP gains, EICS (see Tab.~\ref{tab:EICS and MCP}), and the fragmentation patterns: around 22~\% of the O$_2$ and 2~\% of H$_2$ molecules fragment to their atomic counterparts upon ionization and thus do not appear at 32~u/z and 2~u/z, respectively (see Tab.~\ref{tab:FRAG_and_ISO}). In sum, the measured ratio $r_{meas}$ is biased and has to be corrected in the following way to compute $r_{cor}$: 

\begin{equation}
r_{cor}=\frac{\text{EICS}(\text{O}_2) \cdot \text{FRAG}(\text{O}_2)} {\text{MCP}(\text{H}_2) \cdot \text{EICS}(\text{H}_2) \cdot \text{FRAG}(\text{H}_2)} \, r_{meas} = \frac{2.441 \cdot 0.78} {1.8 \cdot 1.021 \cdot 0.98} \, r_{meas} = 1.057 \, r_{meas}
\label{eq:H2_O2_correction}
\end{equation}

Calculating the ratio of H$_2$/O$_2$ from all experiments in Tab.~\ref{tab:Overview}, excluding those without a significant O$_2$ release and/or those targeting the cooling plate or the Faraday cup, the median ratio and quartiles are:

\begin{equation}
    r_{meas} = 3.6 \pm 1.0
\end{equation}
\begin{equation}
    r_{cor} = 3.8 \pm 1.1
\end{equation}

This average ratio is a factor of two higher than the theoretically expected ratio $r=2.0$. 
However, the actual ratio depends on the electron flux used in a given experiment. For more intense irradiation, the measured H$_2$/O$_2$ ratio approaches asymptotically the theoretical 2:1 ratio. This is illustrated in Fig.~\ref{fig:H2_O2_ratio}: The blue symbols are the additional experiments with a notable H$_2$O release, the orange symbols are those listed in Table~\ref{tab:Overview}. The ratios were corrected for fragmentation, EICS and MCP gains based on Eq.~\ref{eq:H2_O2_correction}. The apparent excess of the H$_2$/O$_2$ release for weak irradiation is a combination of non-saturation, when part of the produced O$_2$ is retained in the ice (assuming all H$_2$ is always released immediately at our ice temperatures), and/or some excess H$_2$ via desorption from other surfaces in the vacuum chamber. This also explains the observation of hydrocarbons, which cannot be produced from pure water ice. Both effects become less relevant for the total ratio at more intense electron beam fluxes where total radiolysis output is increased and saturation is reached, i.e., all freshly produced O$_2$ is released from the ice surface.

\begin{figure}[htp]
  \includegraphics[width=0.95\textwidth]{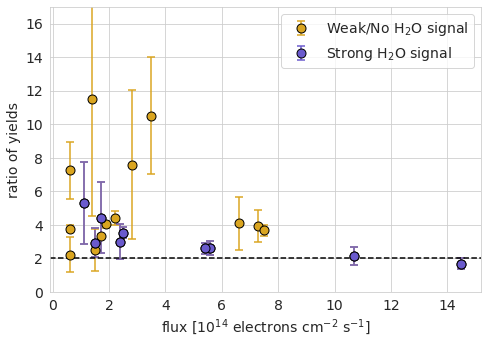}
\caption{Measured steady state ratio (corrected for fragmentation, EICS, and MCP gains) of released H$_2$ versus released O$_2$ as a function of electron fluxes for all experiments with a significant H$_2$ and O$_2$ release.
Orange symbols: experiments with little or no discernible water release (experiments \#1--\#16 without \#3, \#5, and \#9 in Table~\ref{tab:Overview}); blue symbols: experiments with water release $>$ $10^5$~cnts~s$^{-1}$(experiments I--IX in Table~\ref{tab:Overview}). The dotted black line indicates the expected 2:1 ratio once saturation is reached and background contributions become irrelevant.}\label{fig:H2_O2_ratio}
\end{figure}

Apart from the major species H$_2$, O$_2$, and H$_2$O, we also monitored the time series at masses of other radiolysis species found in some previous irradiation experiments (see Section~\ref{sec:introduction}) but were unable to clearly identify H$_3$O, H$_2$O$_2$, or O$_3$. Their relative abundances in the released particle flux with respect to released O$_2$ were $\leq10^{-3}$. This is consistent with the H$_2$O$_2$ concentrations observed in laboratory irradiation experiments by \citeA{Hand2011}. 

We investigated how the H$_2$ and O$_2$ release depend on the characteristics of the electron beam and of the ice sample. In summary, pristine water ice (i.e., not irradiated previously) showed a significantly slower O$_2$ release compared with H$_2$, whereas O$_2$ and H$_2$ decreased equally fast at the end of the electron irradiation, similar to all subsequent rise and fall times of H$_2$ and O$_2$ from pre-irradiated ice. The production rate of the saturated O$_2$ and H$_2$ signals depended linearly on the electron flux. In the following, we first present the timescales of the H$_2$ and O$_2$ release and then show the correlation of the saturated production rates and experiment parameters (electron flux, energy, and sample type).

We quantified the timescale of the rise and decay of the irradiation-induced H$_2$ and O$_2$ release by fitting the time series of the measured signal strengths (in integrated counts/s, without correction for EICS or MCP gains needed) with an exponential function. $\lambda$ is the release constant with corresponding half-life $T_{1/2}=\ln(2)/\lambda$ for the release of radiolytic products:

\begin{equation}
   I(t) = m \cdot \exp(-\lambda \cdot t) + k
   \label{eq: fit}
\end{equation}

The data interval for the fit is chosen from the start and the end of the irradiation until the signal reaches a constant level. An example of the derived fit for both H$_2$ and O$_2$ is shown in Fig.~\ref{fig:fitted_data}. The only species found to deviate from the temporal trend of H$_2$ were O$_2$ and H$_2$O (see Section~\ref{sec:h2o}). 

\begin{figure}[h]
        \includegraphics[width=0.95\textwidth]{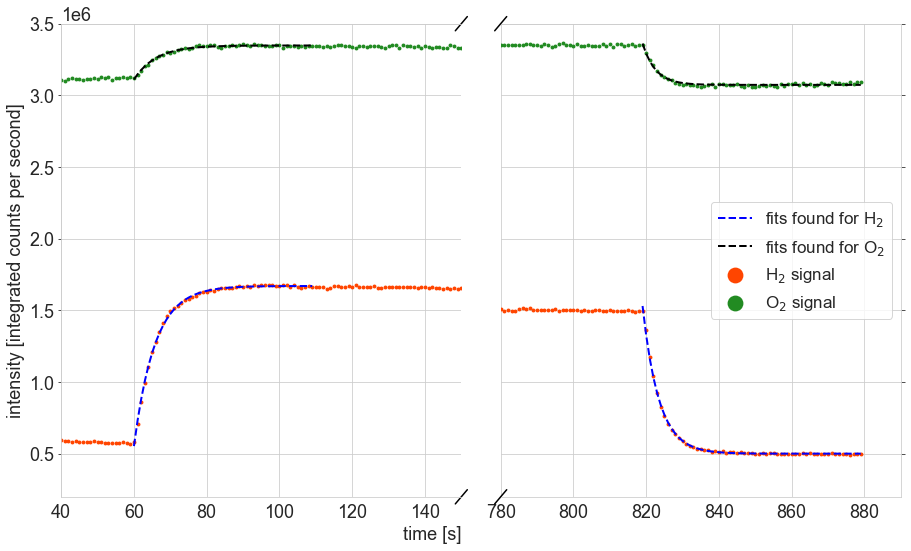}
\caption{Fitting the data with Eq.~\ref{eq: fit} for irradiation \#7 for both the oxygen and hydrogen signal. The irradiation starts at 60~s and ends at 820~s. The data between 140 and 780~s are constant and have been omitted to emphasize the relevant data. The blue fits for the H$_2$, and the black fits for O$_2$ have a coefficient of determination $\text{R}^2$ between 0.97 and 1. The signals show a weak trend of dropping off with ambient pressure over the time of irradiation.}
\label{fig:fitted_data}
\end{figure}

We repeated this fitting for all irradiation experiments with an H$_2$O release $< 10^5$ counts/s and calculated average release constants $\lambda$ for five separate cases: H$_2$ rise, H$_2$ decay, O$_2$ rise for pristine ice, O$_2$ rise for follow-up irradiations, and O$_2$ decay. The results are listed in Table~\ref{tab:Lambdas_summary} and illustrated in Fig.~\ref{fig:Lambdas}. Only $\lambda$-values from a fit with a correlation coefficient $R^2 \geq 0.5$ were used in the averaging. This criterion excluded only 6 out of 68 measured values. Using another cut-off would not significantly change the calculated average half-lives. In Fig.~\ref{fig:Lambdas}, the pristine ice irradiations can be identified by open circles instead of the filled-in circles for the follow-up irradiations. To estimate the errors of the individual data points in Fig.~\ref{fig:Lambdas}, we assumed a normal distribution of the time and intensity error and generated 5,000 additional sets of data with random errors. For each of these new data sets, the half-lives were calculated again. The final error stated in Table~\ref{tab:Lambdas_summary} corresponds to the standard deviation of these half-lives.

\begin{figure}[h]
        \includegraphics[trim=0cm 2cm 0cm 0cm,clip,width=0.99\textwidth]{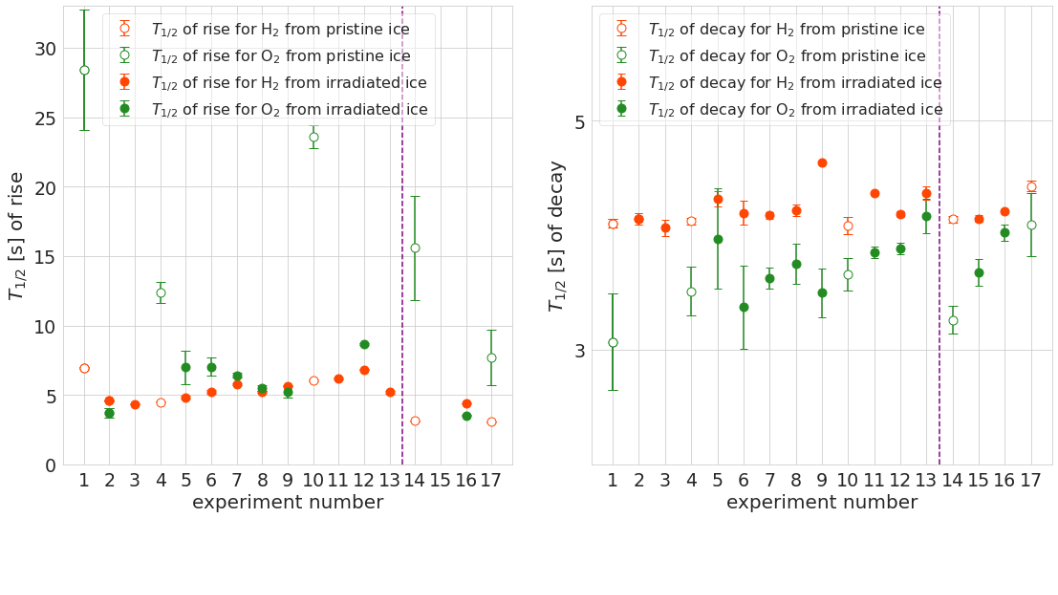}
\caption{Visualization of the half-lives $T_{1/2}$ of molecular oxygen and hydrogen for their rise (left panel) and decay (right panel) for the 17 irradiation experiments defined in Table~\ref{tab:Overview}. The open circles correspond to irradiations of pristine ice. The irradiations on the right of the purple dashed line (\#14-- \#17) were done on coarse-grained ice, the rest of them on fine-grained ice. For the experiment conditions during irradiation see Table~\ref{tab:Overview}.}
\label{fig:Lambdas}
\end{figure}

Figure~\ref{fig:Lambdas} and Table~\ref{tab:Lambdas_summary} show a significant difference between pristine and non-pristine ice regarding the timescale for the rise in O$_2$. The O$_2$ rise for a follow-up irradiation has a half-life of a few seconds, which agrees with the timescale of the H$_2$ rise. All $\lambda$-values in Table~\ref{tab:Lambdas_summary}, with the notable exception of the O$_2$-release from pristine ice, agree within $2\sigma$ with the theoretically expected $\lambda$ for instantaneous changes given the chamber volume and pumping speed of MEFISTO (see Section~\ref{sec:methods}). It takes significantly longer (roughly three times on average) in the case of a pristine ice irradiation (open, green circles in Fig.~\ref{fig:Lambdas}) for the O$_2$ signal to reach a steady intensity. This steady state level is the one stated in Table~\ref{tab:Overview} to be comparable with the H$_2$ release. The most plausible explanation for this delayed release is that O$_2$ is retained in water ice much better than H$_2$, and a longer time is needed before the abundance reaches saturation upon which any additionally produced O$_2$ is being released. Thus, the O$_2$ signal in the vacuum chamber increases as rapidly as for H$_2$ (see $\lambda$-values in Table~\ref{tab:Lambdas_summary}) or other species during subsequent irradiations of a previously irradiated spot of an ice sample. This effect of pre-irradiation was still observed during an irradiation after waiting for 19 hours without intermittent irradiations (irradiation \#5 belonging to initial irradiation \#2 in Table~\ref{tab:Overview}). This agrees with previous observations from water ice films irradiated with electrons \cite{Orlando2003,Petrik2006,Meier2020} and with ions \cite{Reimann1984,Teolis2005}. In our experiments, the half dose or saturation fluence $d_{1/2}$ for pristine fine-grained ice ($T_{1/2}$ from Table~\ref{tab:Lambdas_summary} multiplied with the corresponding electron fluxes from Table~\ref{tab:Overview}) corresponds to $d_{1/2} = 10^{14} \dots 10^{15}$ electrons cm$^{-2}$, in agreement with \citeA{Orlando2003}. This dose seems to be smaller ($d_{1/2}=0.5\times 10^{14}$ electrons cm$^{-2}$) for coarse-grained ice. In Section~\ref{sec:oxygen_storage} we will investigate what the observed release timescales imply for the O$_2$ abundance in icy regolith. 

The different temporal evolution also indicates that the released O$_2$ originated predominantly from the irradiated ice. O$_2$ is not a typical surface contaminant and is not a species outgassing from our electron gun. By contrast, a part of the measured H$_2$ may not have originated from radiolysis in the ice, but is rather a fragmentation product of hydrocarbons from electron gun outgassing given the observed average H$_2$/O$_2$ ratio of $3.8\pm1.1$ compared to the expected ratio of 2.0 (see Section~\ref{sec:relevant_species}). Moreover, having obtained similar rise and decay time constants for H$_2$ and O$_2$ for irradiated ice, i.e., having achieved saturation of radiolytic products in the irradiated ice, indicates that radiolytic H$_2$ and O$_2$ are produced at the rate as theoretically expected. 

\begin{table}[h]
\begin{tabular}{llll}
 & \textbf{$\lambda$ $[$s$^{-1}]$} & \textbf{$\sigma_{\lambda}$ $[$s$^{-1}]$} & \textbf{$T_{1/2}$ [s] } \\
\hline
H$_2$ rise & 0.21 & $\pm$ 0.05 & 3.4 \\
O$_2$ rise, pristine &0.07 & $\pm$ 0.03&  9.8 \\
O$_2$ rise, follow-up & 0.19&  $\pm$ 0.07&  3.6\\
H$_2$ decay & 0.24&  $\pm$ 0.01& 2.9 \\
O$_2$ decay & 0.27 &  $\pm$ 0.05 & 2.5 \\     
\end{tabular}
    \caption{\label{tab:Lambdas_summary} The average release constants $\lambda$ and their corresponding $T_{1/2}$ for the rise and decay of O$_2$ and H$_2$, averaged over all individual experiments of the stated category. The standard deviations $\sigma_{\lambda}$ of the individual $\lambda$-values serve as estimates of the variability over different irradiations.}
\end{table}

 The time series (as seen in Fig.~\ref{fig:timeseries_general}) were used to analyze the steady state release of O$_2$ and H$_2$: a linear fit of the species-specific time series before and after the irradiation was subtracted from the intensity during irradiation. This is possible as, in all irradiations, the signal decays fast enough back to initial levels after 10 seconds. 
The H$_2$ and O$_2$ production as a function of the electron flux (see Fig.~\ref{fig: prod vs current}) shows the expected linear correlation with the electron flux. 
The observed production rates (i.e., the count rates integrated over the peak width of a species listed in Table~\ref{tab:Overview}) were divided by the irradiated surface area of the electron beam also used to convert electron rates into fluxes. The numbers were not corrected for EICS, MCP gains, or H$_2$/O$_2$ excess. Linear regressions are plotted as dashed lines. At low electron fluxes, the scatter of production for a given flux is considerable, but no systematic deviations are apparent with two exceptions: The irradiation of the Faraday-cup produces no significant O$_2$ release and coarse-grained ice seems to release more O$_2$ and H$_2$ than fine-grained ice for similar electron fluxes (compare open and filled symbols in Fig.~\ref{fig: prod vs current}). The latter trend could be due to the different thermal properties of the ice types (see Section~\ref{sec:h2o}) as the yields for H$_2$ and O$_2$ release are known to increase with ice temperatures (see Section~\ref{sec:introduction}).

\begin{figure}[h]
  \includegraphics[width=0.95\textwidth]{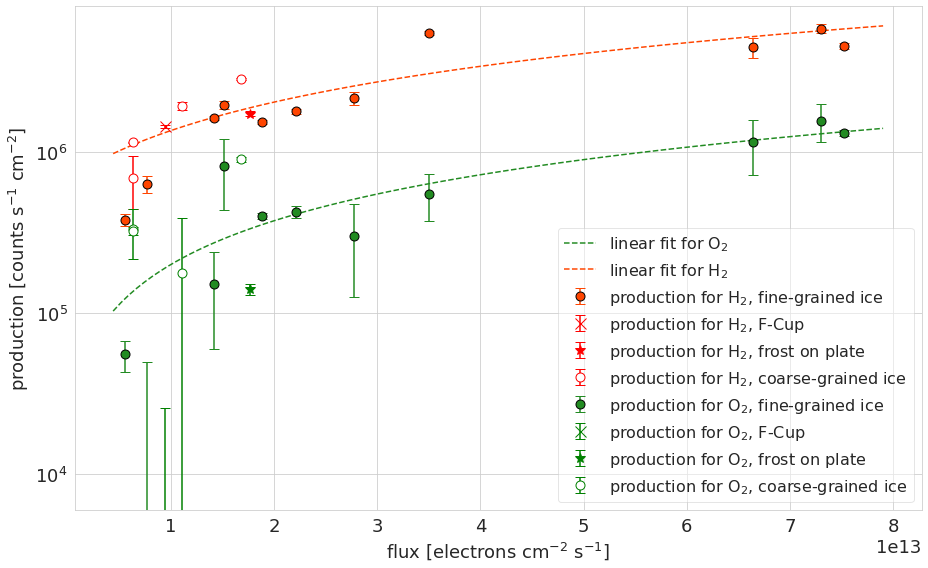}
\caption{H$_2$ (red symbols) and O$_2$ (green symbols) steady state production in integrated and averaged counts per second and cm$^2$ compared to the electron flux. The linear regressions were determined using the least squares method and using the errors to weigh them. The errors correspond to the standard deviation of the signal during irradiation. 
The irradiations of fine-grained ice are displayed in filled circles and the ones of coarse-grained ice in open circles. The ``X'' symbols indicates the result of the Faraday cup irradiation, which resulted in an H$_2$ signal, but no O$_2$ release significantly higher than the background level. The asterisks denote the irradiation of water frost on the cooling plate.} \label{fig: prod vs current}
\end{figure}

We also examined if O$_2$ or H$_2$ production depended on electron energy. Whereas the H$_2$ production was not affected by electron energy, the O$_2$ production (divided by flux because of the linear relation between production and flux) was found to decrease on average by a factor of two when the energy increased from 0.5 to 3 keV. This applies for all irradiations \#1 to \#17 without the Faraday cup irradiation and also for the subset including only fine-grained ice. However, this decrease is not statistically significant because of the available data points and potential cross-correlation effects from pristine and pre-irradiated ice.

\section{Discussion}\label{sec:discussions}

\subsection{Production and Retention of O$_2$ and H$_2$ in the Ice}

Production of O$_2$ was found during nearly all irradiations and was observed at energies as low as 0.5~keV and fluxes of roughly $10^{13}$ electrons cm$^{-2}$ s$^{-1}$. The production follows a nearly linear relation with the flux (Fig.~\ref{fig: prod vs current}). This is what is to be expected according to the calculations performed by \citeA{Petrik2006}. They derived that the measured O$_2$ production rate should be proportional to the concentration of excitation products generated by the incident electron beam. 

The dependence of the O$_2$ yield on the energy of the electrons is more complicated: \citeA{Teolis2017} emphasize that while many experiments have confirmed the proportionality below 100 eV, almost no experiments with higher energies were performed. In the low energy regime, the linearity was confirmed by both \citeA{Orlando2003} and \citeA{Petrik2006}, leading the latter to conclude that the first step of O$_2$ production is ionization. \citeA{Teolis2017} predict a fall off at around 0.4~keV as the electrons penetrate to depths where the energy delivered to the 30~$\text{\AA}$ surface layer (and therefore available for O$_2$ formation) declines with increasing electron energy. This has been tested with higher energetic electrons impacting compact ice films on microbalances \cite{Meier2020} who showed the predicted decrease in yield with rising energies. \citeA{Galli2018}, on the other hand, found no significant change of O$_2$ yields from granular ice samples with electron energies between 0.5 and 10 keV. In this study, we observed no clear trend of H$_2$ production with energy, whereas the O$_2$ production decreased by a factor of two from 0.5 to 3 keV. This trend was not significant and needs to be studied in more detail in the future for enough cases with different types of porous ice.

When discussing the retention of O$_2$ in the ice, previous experiments with water ice films also found a slower O$_2$ release for irradiations of pristine ice \cite{Petrik2006,Meier2020}. In the irradiations of pristine ice films performed by \citeA{Petrik2006} at $T$ = 80 K, the O$_2$ production took around 5--15 seconds until reaching steady state. They also stated that the production for $T > 80$~K increases with $1- e^{-\alpha t}$, which we can confirm, but do not mention any values for $\alpha$. In their follow-up irradiations, this timescale was reduced. To our knowledge, previous studies did not state explicit time constants and derive corresponding O$_2$/H$_2$O ratios in the ice (see Section~\ref{sec:oxygen_storage}).  

\citeA{Meier2020} discovered that a follow-up irradiation could produce up to six times more O$_2$ than the same irradiation on pristine ice if the first irradiation were done at higher energies than the follow-up. This could not be tested with our data, as we varied the experimental parameters (energy, current, ice type). This will be the subject of further study in our next irradiation experiments. 
It is of interest for how long the additionally produced O$_2$ can be stored in the ice \cite<respectively its O$_2$H precursor according to>[]{Petrik2006}. 
We verified that follow-up irradiation up to 19 hours after the stop of the initial irradiation (having kept the ice sample at temperatures below 120 K in the meanwhile) showed a release of O$_2$ from the ice as rapidly as all other species, i.e., the oxygen abundance in the previously irradiated ice must have remained high for at least 19 hours. 
Few other studies mention time scales, i.e., how long they waited until re-irradiating the pre-irradiated ice (``few minutes'' or ``several minutes'' according to \citeA{Boring1983, Reimann1984, Teolis2005}).
More experiments at various temperatures with longer resting phases between irradiations and constant monitoring of any O$_2$ release are therefore needed to investigate how long radiolysis oxygen remains trapped in water ice.
Regarding H$_{2}$, it is well known that it cannot remain trapped at these experiment temperatures \cite{Bar-Nun1985}. Our experiments confirmed this; the timescales of rise do not differ for pristine and follow-up irradiations, nor do the total yields. 

The H$_2$/O$_2$ ratio $>2$ at low electron fluxes (see Fig.~\ref{fig:H2_O2_ratio}) can be attributed to the additionally released H$_2$ from surfaces and the electron gun or to O$_2$ first having to accumulate to a certain abundance in the ice before release. It is noteworthy in this context that \citeA{Grieves2005} found a much higher ratio of released D$_2$ compared with O$_2$ below 100 K when they warmed up D$_2$O ice pre-irradiated with electrons. \citeA{Grieves2005} found that O$_2$ was not released in the same temperature range as D$_2$, but was retained within the ice until water ice sublimation temperatures (140 K) were reached. The authors suggested that O$_2$ may only reach a macropore that is in the immediate vicinity because it does not effectively move through microporous channels. Likewise, \citeA{Zheng2006} saw an O$_2$ signal when irradiating at 1000~nA and 5~keV, which was only 4.6~\% compared to their H$_2$ signal. That leads us to the question if the H$_2$/O$_2$ excess (3.8 instead of 2.0 on average) was only due to surface contamination or if it was also caused by more effective trapping of O$_2$ molecules in the ice so long as the temperature remained below 140~K. This would explain why the H$_2$/O$_2$ ratio approached 2.0 for intense electron fluxes where the irradiated volume of the ice samples probably were warmed up and started sublimating.

\subsection{Implications for Icy Satellites and Icy Grains}
\label{sec:oxygen_storage}

In Section~\ref{sec:timescales}, we studied the different timescales for the release of O$_2$ upon either irradiation of pristine ice or follow-up irradiation. The timescale of the rise of O$_2$ can be converted into a ratio of O$_{2}$/H$_{2}$O in the irradiated ice layer. For that, $\lambda=0.113\pm 0.090$ s$^{-1}$ (respectively $T_{1/2} =6.11 \pm 4.84$ s) of the difference between the O$_{2}$ timescale of release from pristine versus pre-irradiated ice is used, taking into account the chamber volume and pumping speed of MEFISTO (see Table~\ref{tab:Lambdas_summary}). The ratio of O$_2$/H$_2$O can be calculated by dividing the number of retained O$_{2}$ by the number of H$_{2}$O molecules in the irradiated layer of thickness $d$ (penetration depth):

\begin{equation}
r(\textup{O$_2$/H$_2$O}) = \frac{{Y_{O2}} \, j_e\, m_{\textup{mol}}} {A \, d \, \rho \, N_\textup{A} \, q_e \, \lambda} \approx 0.004  \label{eq:ratio} 
\end{equation}

For this estimate, we assumed the averages for the cases at 1~keV energy with the electron sputtering yield of O$_2$ $Y_{O2} = 1.5$ \cite{Galli2018}, $j_e$ = 3 $\mu$A, $d$ = 46~nm,  the corresponding penetration depth of 1~keV electrons,  $m_{\textup{mol}}=18$ g mol$^{-1}$, $N_A$ Avogadro's constant, and $\rho=1$ g cm$^{-3}$ (for a lower density, the penetration depth would increase accordingly) and $A= 0.4$ cm$^2$. 
The uncertainty of this result is at least as large as the uncertainty of the time constant $\lambda= 0.113\pm 0.090$ s$^{-1}$, corresponding to a ratio of O$_2$/H$_2$O of (0.2--2.0)$\cdot 10^{-2}$.
This implies that molecular oxygen can be formed and retained in 100~K water ice at a ratio of roughly $10^{-2}$ O$_{2}$ to H$_{2}$O when the water is irradiated by electrons that are readily available in both flux and energy in the vicinity of icy moons inside the magnetospheres of Jupiter and Saturn. The O$_{2}$ to H$_{2}$O ratio derived from our experiments is of the same order of magnitude as the one inferred from observations of the Galilean satellites (see Section~\ref{sec:introduction}). 

\section{Conclusions}\label{sec:conclusions}

The experiments presented here demonstrated sputtering and radiolysis from water ice regolith irradiated by electrons at conditions representative for the icy moons of Jupiter. The experiment setup allowed us to identify the species released from the ice samples and to analyze their temporal evolution. We also discussed the technical challenges of the current setup, namely a potential cold-trap effect and hydrocarbon contaminants contributing to some signals in the mass spectrometer. Future experiments need to quantify cold-trapping by varying the observation geometry, while the latter challenge requires a lower partial pressure of carbon-bearing molecules including CO$_2$ inside MEFISTO.

The radiolysis of H$_{2}$O into H$_{2}$ and O$_{2}$ upon electron irradiation was the dominant reaction for the types of water ice samples investigated, and H$_{2}$ and O$_{2}$ could be detected for almost every irradiation experiment. Minor H$_2$O contributions were measured sometimes, but the H$_2$O sputtered or desorbed from the ice may be of the same order of magnitude as the released radiolysis products H$_{2}$ and O$_{2}$, considering the cold-trap effect that biases against H$_{2}$O detection.

We confirmed previous observations made with water ice films that pristine ice exhibits unique O$_{2}$ release and saturation properties. We were able to measure the saturation timescale and verified that O$_{2}$ from previously saturated ice samples was released within seconds upon follow-up irradiations. Oxygen can be stored for long times, and the effect of saturation was observed even when there were hours between pristine and follow-up irradiation. The O$_{2}$/H$_{2}$O ratio derived from our experiments is $\approx$ 0.01 in the irradiated ice.

Other possible radiolysis species such as H$_{2}$O$_{2}$, O$_{3}$, were not detected yet. An important issue for the detection is that the fragmentation patterns are specific to the used mass spectrometer \cite{Gasc2015}. To detect the water-related species that share a similar mass with fragmentation products of water, more accurate fragmentation patterns have to be measured directly with our setup to an accuracy better than 1\%. Otherwise, the release of water must be inhibited (by choosing low energies and flux) to detect H$_2$O$_2$ and O$_3$. The non-detection of H$_2$O$_2$ is understandable given the current detection limit and the ice temperatures that are warmer than those where irradiation-induced H$_2$O$_2$ production was previously observed to be efficient \cite{Zheng2006,Hand2011}.

When irradiating the ice samples with higher energies and electron flux, we found sublimation effects of water ice: In these cases, the measured H$_{2}$O reached or surpassed the measured H$_{2}$ and O$_{2}$. This release was to be expected for high electron fluxes. Interestingly, the release was found to differ between two ice types with different grain size and porosity. This can be explained by the different thermal properties of the ice types, but additional laboratory experiments are required to provide more data on different ice types irradiated under the same conditions.

In summary, irradiation of water ice with electrons (and other energetic particles) is a plausible explanation for the occurrence of the O$_{2}$ observed on the surface of icy moons. For comets, the observed O$_{2}$/H$_{2}$O ratio in the gas outflow of 67P/C-G of (3.80 $\pm$ 0.85)$\cdot 10^{-2}$ \cite{Bieler2015} is also consistent with the ratio of (0.2--2.0)$\cdot 10^{-2}$ in the irradiated surface derived from our experiments in the laboratory. However, in the case of comets, the radiolysis hypothesis faces the challenges that the most of the surface is covered by a non-ice crust and that the oxygen isotope ratios in O$_{2}$ and H$_{2}$O do not match \cite{Mousis2016,Altwegg2020}: any O$_{2}$ created by radiolysis when the ice was last exposed to the space environment has to be preserved in the interior for billions of years until the comet approaches the sun and becomes active. 

\section*{Open Research}
The original mass spectra obtained  for this study and the software (Jupyter notebooks) to create figures for this publication are openly available via a Zenodo repository \cite{Galli2024}. We invite the reader to test our software and are happy for every feedback.

\begin{acknowledgments}
Acknowledgments: The authors wish to acknowledge support from the Swiss National Science Foundation (SNSF; 200020\_207409). Work by M.R. was funded by the Canton of Bern and the SNSF (200020\_207312). The authors are also very grateful for the long-term support provided by laboratory technicians and engineers Joël Gonseth, Raphael Hänggi, Martin-Diego Busch, Daniele Piazza, and Adrian Etter. 

\end{acknowledgments}

\bibliography{submission.bib}

\end{document}